\documentstyle[twoside,fleqn,epsf,amssymb,espcrc2]{article}
\begin{document}

\title{I=2 pion scattering length with the parametrized fixed point action.}

\author{K.~Jimmy Juge [BGR (Bern-Graz-Regensburg) Collaboration]}

\begin{abstract}
We report on the $\pi\pi$ scattering length in the $I=2$ channel using the parametrized fixed point (p-FP) action. Pion masses of 320 MeV were reached in this quenched calculation of the scattering length.
\end{abstract}

\maketitle

\section{Introduction}
The pion scattering length is an important observable characterizing dynamical effects of the strong interaction and its ab initio calculation on the lattice is an important nonperturbative test of QCD. In the literature several attempts to compute the scattering length with Wilson fermions \cite{Sharpe,Fukugita,Liu,JLQCD,CPPACS} and staggered fermions \cite{Sharpe,Fukugita} can be found. However, calculations with a chiral Dirac operator which allows for a better control of the small mass region are still missing. In full QCD the scattering length is a quantity which vanishes in the chiral limit, while it is power divergent in the quenched theory \cite{Bernard} and eventually a study in the full theory is desirable. In this contribution, we present results of the scattering length calculated using the p-FP action. (Calculations with chirally improved fermions is also underway.) The chiral and scaling properties of the light hadron spectrum using p-FP and chirally improved actions were reported by the collaboration in an earlier publication \cite{spectrum}. 
\section{Method and simulation details}
The $\pi^+\pi^+$ scattering length is extracted using the standard finite volume technique of L\"uscher \cite{Luscher}. The energy shift of the two pion state, $\delta E(L)$, in a finite volume is related to the scattering length, $a_0$, in the infinite volume limit through the relation, 
$$\delta E=-\frac{4\pi a_0}{M_\pi L^3}\{1+c_1\frac{a_0}{L}+c_2\frac{a_0^2}{L^2}\}+\dots$$
where $c_1 = -2.837297$ and $c_2 = 6.375183$. 

The two pion state is created using a Gaussian smeared source for the quark propagator on a single timeslice. A second smeared source which was displaced in time by one lattice unit was used on small lattices to check for the systematics and the often claimed complications due to Fierz mixing. Point sink operators were used on a single timeslice regardless of the position of the sources.

Cutoff effects were studied in a small volume $(1.2\ \rm{fm})^3$ using three different lattice spacings ($0.08, 0.10, 0.15 \rm{fm}$) and in an intermediate volume of $(1.8\ \rm{fm})^3$ with $\rm{a}\sim0.10$ and $0.15$ fm. A wide range of quark masses were studied, with the smallest quark mass resulting in $M_\pi/M_\rho\sim0.35$. In this quenched study, we keep those masses with $M_\pi L\gtrsim4$. 
\begin{figure}
\vspace{-2ex}
\begin{center}
\epsfxsize=2.5in \epsfbox{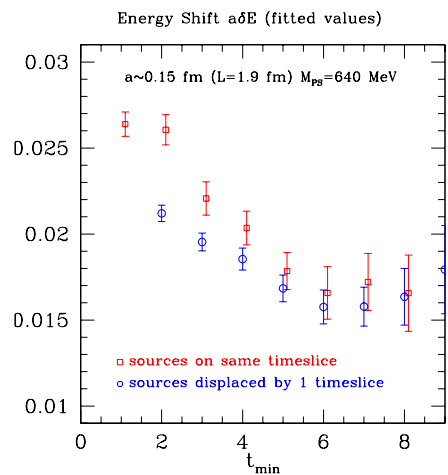}
\end{center}
\vspace{-8ex}
\caption[]{Fitted values of the energy shift as a function of $t_{\rm{min}}$ for the two sources on the same timeslice and displaced by one unit.}
\label{fierz}
\vspace{-2ex}
\end{figure}
\begin{figure}
\begin{center}
\epsfxsize=2.5in \epsfbox{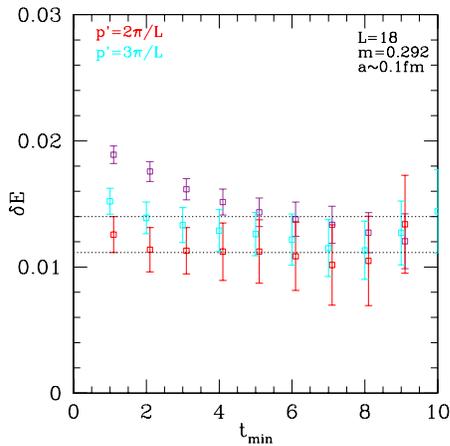}
\end{center}
\vspace{-8ex}
\caption[]{Fit results using different values for the energy of the excited state. The smaller error results are from a single exponential fit.}
\label{excited}
\end{figure}
\section{Analysis}
We extract the energy shift of the two pion state in a finite volume by taking ratios of two pion correlation functions and single pion correlation functions. We follow the procedure and notation of Ref.~\cite{Sharpe} and construct the ``direct'' and ``crossed'' contractions of
$$<\sum_{\vec{x}_1}{\mathcal O}(\vec{x}_1,t)\sum_{\vec{x}_2}{\mathcal O}(\vec{x}_2,t){\mathcal S}_2{\mathcal S}_1>$$
where ${\mathcal O}$ annihilates a $\pi^+$ state and ${\mathcal S}_i$ are the Gaussian sources for the pions and divide by the two single pion propagators. The large time behaviour of the ratio is given by $Ze^{-\delta Et}$ in the full theory.
\subsection{Fitting procedure}
Periodic boundary conditions were used in the time direction so that we must take into account a backwards propagating pion. The fitting form of the  correlation function was
$$Z\frac{e^{-(2M_\pi+\delta E) t}+e^{-(2M_\pi+\delta E)(T-t)}+ze^{-M_\pi T}}{e^{-2M_\pi t}+e^{-2M_\pi (T-t)+2e^{-M_\pi T}}}.$$
The pion mass was fitted from the pion propagator alone within the same bootstrap sample.
The momentum spacings of the pions in the largest volume is quite small which resulted in significant excited state contamination (higher momenta states). In this case, we fit to a sum of exponentials (assuming this is the case) to take this into account. Only the amplitude of the excited state was left as an extra fitting parameter and we vary the energy of this state over a wide range to monitor the effect of the choice of this energy. The results are shown in Fig.~\ref{excited}. It can be seen in this example that the ground state energy is nearly independent of the value of the energy of the excited state, however, with much larger errors than with a single exponential fit. 
\section{Results/Discussion}
\begin{figure}
\vspace{-2ex}
\begin{center}
\epsfxsize=2.5in \epsfbox{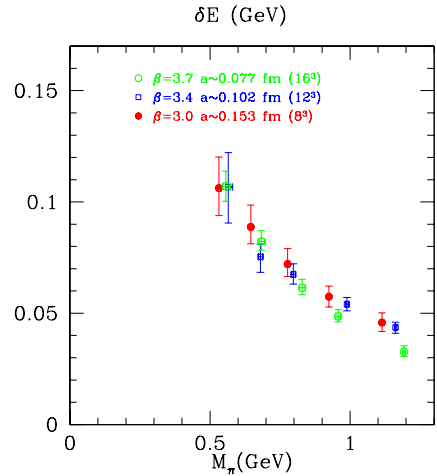}
\end{center}
\vspace{-8ex}
\caption[]{$\delta E$ for three different lattice spacings in a small volume of $(1.2\ \rm{fm})^3$. Very little or no cutoff effects are observed for $M_\pi<1\  \rm{GeV}$. We choose $\delta E$ here because the volume is quite small.}
\label{cutoffv}
\vspace{-2ex}
\end{figure}
The effect of displacing the second source by one timeslice is shown in Fig.~\ref{fierz}. The asymptotic value agrees within errors but there is less excited state contamination when the two pions are not created on the same timeslice. The effect of the second source is being investigated on the largest lattice where this contamination is the source of the large errorbars as was mentioned in the previous section.
 
\begin{figure}
\vspace{-2ex}
\begin{center}
\epsfxsize=2.5in \epsfbox{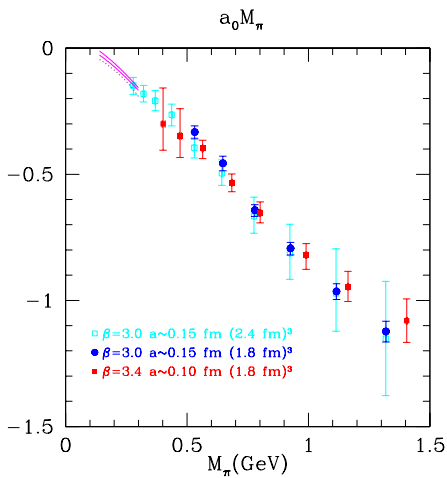}
\end{center}
\vspace{-8ex}
\caption[]{$a_0M_\pi$ for two different lattice spacings in an intermediate volume of $(1.8\ \rm{fm})^3$. A large volume result is also shown for reference. The curves are the perturbative results.}
\label{cutoffV}
\vspace{-2ex}
\end{figure}

The cutoff effects of $\delta E$ in the small volume, $(1.2\ \rm{fm})^3$, is shown in Fig.~\ref{cutoffv}. We observe no significant lattice artifacts for pseudo-scalars lighter than $\sim1$ GeV. A similar analysis in the large volume for $a_0M_\pi$ (Fig.~\ref{cutoffV}) confirms that the lattice artifacts are well under control and that the results from $a\sim0.15\ \rm{fm}$ and $L=2.4\ \rm{fm}$ are very close to the continuum and infinite volume limit.

\begin{figure}
\begin{center}
\leavevmode
\vspace{-7ex}
\epsfxsize=3.1in\epsfysize=3.0in\epsfbox{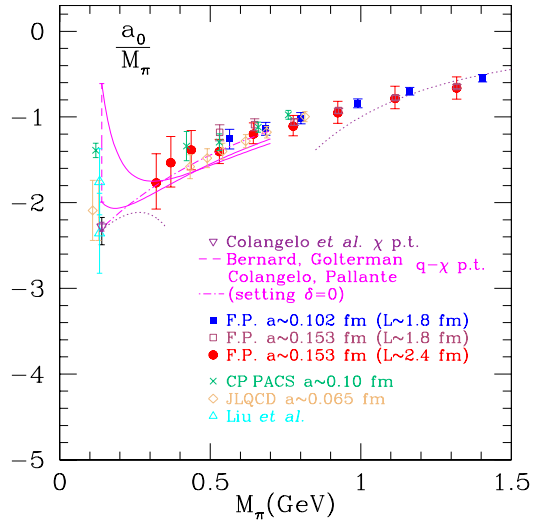}
\end{center}
\vspace{-8ex}
\caption[]{Summary of our results and other recent calculations. The spread in the quenched chiral p.t.~calculation corresponds to the range $0.10<\delta<0.20$. $\delta=0$ case is also shown where the mass dependence comes solely from $F_\pi$. The dashed line at large masses is $-1/M_\pi^2$.
}
\label{summary}
\end{figure}

In Fig.~\ref{summary}, we plot $a_0/M_\pi$, against $M_\pi$ in physical units. We also include in this plot some recent results from other groups as well as quenched chiral perturbation theory \cite{Bernard} and chiral perturbation theory predictions \cite{Colangelo}. The mass range of the perturbative calculations shown in this figure were chosen for illustration and does not suggest convergence of the series. The effect of the $\delta^2/M_\pi^2$ term in the quenched theory is still not entirely visible at the present quark masses. The results also suggest that pseudo-scalars with masses less than $300\ \rm{MeV}$ may be needed for a reliable extrapolation to the physical pion mass. It remains a challenge to perform a full QCD simulation with such light quarks.
\section*{ACKNOWLEDGEMENTS}
K.J. would like to thank G.~Colangelo and J.~Gasser for providing details of their calculations and insightful discussions. The simulations were done on the Hitachi SR8000 at the Leibniz Rechenzentrum in Munich and at the Swiss Center for Scientific Computing in Manno. This work was partially supported by the European Community's Human Potential Programme under contract HPRN-CT-2000-00145.


\begin{thebibliography}{9}
\bibitem{Sharpe}
   R.~Gupta {\it et al.}, Nucl.\ Phys.\ {\bf B383} (1992) 309.
\bibitem{Fukugita}
   M.~Fukugita {\it et al.}, Phys.\ Rev.\ {\bf D52} (1995) 3003.
\bibitem{Liu}
   C.~Liu {\it et al.}, Nucl.\ Phys.\ {\bf B624} (2002) 360.
\bibitem{JLQCD}
   JLQCD Collaboration, Phys.\ Rev.\ {\bf D66} (2002) 077501.
\bibitem{CPPACS}
   CP-PACS Collaboration, Phys.\ Rev.\ {\bf D67} (2003) 014502.
\bibitem{Bernard}
   C.~Bernard and M.~Golterman, Phys.\ Rev.\ {\bf D53} (1996) 476; G.~Colangelo and E.~Pallante, Nucl.\ Phys.\ {\bf B520} (1998) 433.
\bibitem{spectrum}
   C.~Gattringer {\it et al.}, hep-lat/0307013.
\bibitem{Luscher}
   M.~L\"uscher, Commun.\ Math.\ Phys. {\bf 104} (1986) 177; {\bf 105} (1986) 153; Nucl.\ Phys.\ {\bf B354} (1991) 531.
\bibitem{Colangelo}
   G.~Colangelo {\it et al.}, Nucl.\ Phys.\ {\bf B603} (2001) 125.
\end{thebibliography}
\end{document}